\documentclass[prc,aps,superscriptaddress,showpacs,nofootinbib,twocolumn]{revtex4}
\usepackage{amssymb}
\usepackage[dvips]{graphicx}
\usepackage[english]{babel}
\usepackage{indentfirst}
\usepackage{amsxtra}
\usepackage{amsmath}
\usepackage{supertabular}
\usepackage{multirow}
\usepackage[mathcal]{eucal}
\usepackage[usenames]{color}
\usepackage{ulem}

\begin{document}
\title {\bf Magicity of the $^{52}$Ca and $^{54}$Ca isotopes and tensor contribution within a mean--field approach}

\author{Marcella Grasso}
\affiliation{Institut de Physique Nucl\'eaire, IN2P3-CNRS, Universit\'e Paris-Sud, 
F-91406 Orsay Cedex, France}

\begin{abstract} 
We 
investigate the magicity of the isotopes $^{52}$Ca and $^{54}$Ca, that was recently confirmed by two 
experimental measurements, and relate it to like--particle and neutron--proton tensor effects within a mean--field description. 
By analyzing Ca isotopes, 
we show  
that the like--particle tensor contribution induces shell effects that render these nuclei more magic than 
they would be predicted by neglecting it. 
In particular, such induced shell effects are stronger in the nucleus $^{52}$Ca and the single--particle 
gaps are increased in both isotopes due to the tensor force. 
By studying $N=32$ and $N=34$ isotones, neutron--proton tensor effects may be isolated and their role analyzed. It is shown that neutron--proton tensor effects lead to increasing $N=32$ and $N=34$  gaps, when going along isotonic chains, from $^{58}$Fe to $^{52}$Ca, and from $^{60}$Fe to $^{54}$Ca, respectively. 

The mean--field calculations are perfomed by employing one Skyrme parameter set, that was 
introduced in a previous work by fitting the tensor parameters together with the spin--orbit strength. 
The signs and the values of the tensor strengths are thus checked within this specific application. 
The obtained results indicate that the employed parameter set, even if generated with a 
partial adjustment of the parameters of the force, leads to the correct shell  behavior and provides, in particular, a description of the magicity of $^{52}$Ca and $^{54}$Ca within a pure mean--field 
picture with the effective two--body Skyrme interaction. 
\end{abstract} 

\vskip 0.5cm \pacs {21.60.Jz, 21.10.-k,21.10.Pc} \maketitle 
% 

%-----------------------------------------------------------------------
\section{Introduction}
We recently introduced tensor parametrizations for the phenomenological effective forces Skyrme and Gogny, 
by taking into account a tensor force of zero and finite range, respectively \cite{grasso2013}. The tensor 
parameters were adjusted on top of already existing Skyrme and Gogny parametrizations, by employing a 
three-step fitting procedure inspired by Refs. \cite{zale1,zale2} and by modifying also, simultaneously, 
the spin--orbit strength.  
This work was done as an exploratory study to identify the correct signs and regions for the values of 
the tensor parameters, and can be viewed as a preparatory study for a global fit of all the parameters, 
especially in the Gogny case, where much less work has been done including the tensor force. 

In the present work, we employ one of the Skyrme parametrizations introduced in Ref. \cite{grasso2013} and show that (even if it was not found with an adjustment of all 
the parameters of the force) it properly accounts, within a mean--field picture, for shell effects that were recently confirmed by 
experimental measurements, namely the magicity of the calcium isotopes $^{52}$Ca and $^{54}$Ca. 

Experimental studies strongly indicate $N=32$ as a new magic number in Ca isotopes due to the high 
energy of the first 2$^+$ state in this nucleus \cite{huck,gade}. More recently, 
high--precision mass measurements were performed for the neutron--rich Ca isotopes $^{53}$Ca and $^{54}$Ca by 
employing the mass spectrometer of ISOLTRAP at 
CERN \cite{wienholtz}. The found results and, in particular, the trend obtained for the two--neutron separation energies $S_{2n}$, 
definitely confirmed the magicity of the nucleus $^{52}$Ca.
For the nucleus $^{54}$Ca, the first experimental spectroscopic study on low--lying states was performed very recently 
with proton knockout 
reactions at RIKEN \cite{riken}. This study (in particular, the high energy of the first $2^+$ state) provided a robust experimental signature indicating the magic nature of the nucleus $^{54}$Ca. 
Some comparisons with theoretical calculations have been included in the two experimental studies of Refs. \cite{wienholtz,riken}. 
In Ref. \cite{wienholtz}, the experimental $S_{2n}$ energies have been compared with the results of microscopic calculations based on chiral interactions, with coupled--cluster and shell--model results, as well as with Energy Density Functional (EDF) results based on the mean--field approximation. In Ref. \cite{riken}, shell--model results \cite{honma} have been compared with the experimental data and the important role played by the neutron--proton (n--p) tensor contribution has been investigated. It was also mentioned in Ref. \cite{riken} that  recent calculations including three--body forces \cite{holt,hagen} provide a very good agreement with the experimental results. It was stressed that, in the shell--model calculations of Ref. \cite{honma}, the effect of three--body forces is included empirically. This explains why the obtained  results are very similar to those of Refs. \cite{holt,hagen}, where the three-body contribution is taken into account. 

In this work, we perform Hartree-Fock calculations in spherical symmetry and neglect pairing correlations. The calculations are done in coordinate space. We do not need more sophisticated models because our objective is to isolate the genuine tensor contribution: 
We show that the tensor force, and its induced like--particle and n--p effects, may describe the magicity of $^{52}$Ca and $^{54}$Ca within a simple mean--field scheme.
   
By analyzing Ca isotopes, 
it is shown in particular that an enhancement of the magicity of $^{52}$Ca and $^{54}$Ca may be 
predicted within mean--field calculations by including the like--particle contribution generated by the 
tensor force. On the other side, 
by analyzing $N=32$ and $N=34$ isotones, the role played by the n--p tensor contribution is  investigated. 
One of the Skyrme parametrizations introduced in Ref. \cite{grasso2013} is employed, that was  
 constructed on top of the 
SLy5 \cite{sly5} Skyrme force.  
In the three--step fitting procedure adopted in Ref. \cite{grasso2013}, the first adjustment was done 
 to tune the spin--orbit strength (before tuning the tensor 
parameters) to reproduce the neutron $f$ spin--orbit splitting in the nucleus $^{40}$Ca. This nucleus 
is spin saturated and, consequently, the tensor force does not have any effect on its spectroscopic 
properties. After this first adjustment, the tensor parameters were tuned to reproduce the neutron $f$ 
spin--orbit splitting first in the nucleus $^{48}$Ca and then in the nucleus $^{56}$Ni. 
Details about this procedure may be found in Ref. \cite{grasso2013}. 
The difference of our Skyrme parameter set with respect to the tensor parametrization published by 
Col\`o {\it et al.} in Ref. \cite{colo} (also introduced on top of SLy5) is that the spin--orbit 
strength is simultaneously modified in our case.     
For our Skyrme set, the spin--orbit strength was reduced to 101 MeV fm$^5$,  
with respect to the value in the original force, and the parameters responsible for the like--particle and 
the n--p tensor effect were adjusted to the values $\alpha_{\rm T} = -170$ and 
$\beta_{\rm T} = 122$ MeV fm$^5$, respectively. 
The parameters $\alpha_T$ and $\beta_T$ are related to the parameters $U$ and $T$ as follows, 
\begin{eqnarray}
\nonumber
\alpha_{\rm T}&=&\frac{5}{12} \, U, \\
\beta_{\rm T}&=&\frac{5}{24} \left ( T+U \right ) ,
\end{eqnarray}
where $T$ and $U$ are the strengths of the Skyrme zero--range tensor force in even and odd states of relative motion, 
respectively \cite{skyrme}. 

The article is organized as follows. In Sec.~\ref{like} we analyze the enhancement of magicity in the 
isotopes $^{52}$Ca and $^{54}$Ca, that is  related to the like--particle tensor contribution. 
In Sec.~\ref{np}, $N=32$ and $N=34$ isotones are analyzed and the n--p tensor effects are investigated. 
In Sec.~\ref{masses} the two--neutron separation energies are compared with the experimental data for the nuclei $^{50}$Ca, $^{52}$Ca, and $^{54}$Ca. 
 In Sec.~\ref{concl} conclusions are drawn. 

\section{Magicity of the isotopes $^{52}Ca$ and $^{54}Ca$. Like--particle tensor effects}
\label{like}

We evaluate the single--particle neutron gap for four Ca isotopes, the closed--shell nuclei $^{40}$Ca and $^{48}$Ca  
and the two systems that we wish to analyze here, $^{52}$Ca and $^{54}$Ca. Ca isotopes are 
spin saturated in protons. This means that tensor effects on neutron single--particle states may be 
induced in practice only  by the like--particle neutron--neutron (n--n) contribution along the isotopic chain. We show in Fig. 1 the 
single--particle gaps obtained for the four isotopes with our effective Skyrme force by quenching the 
like--particle tensor strength (dashed line) and by switching it on (solid line). 
The reported gaps refer to $N=20$, $N=28$, $N=32$, and $N=34$ for $^{40}$Ca, $^{48}$Ca, $^{52}$Ca, and 
$^{54}$Ca, respectively. 
Two effects may be 
observed. First, a global enhancement of the single--particle gaps is visible due to the tensor contribution. 
This enhancement is found also for the closed--shell nucleus $^{48}$Ca, where the gap is increased by 
1.3 MeV. The corresponding experimental value is 5.4 MeV \cite{volya}. 
This means that the inclusion of the n--n tensor contribution leads to the correct shift, towards the 
experimental result. 
For the nucleus $^{40}$Ca, the results obtained with and without the n--n tensor effect are obviously almost the same, since this nucleus is fully spin saturated and the tensor force is not active there. 
We notice that the gap variation due to the inclusion of the n--n tensor contribution is more important for $^{52}$Ca than for $^{48}$Ca and $^{54}$Ca, as will be explained below. 
In the figure, also the results obtained with the original SLy5 force are reported (blue dotted line). We may observe that our new set of parameters provides a global improvement of the results: it can be seen that both single--particle gaps obtained for $^{40}$Ca and $^{48}$Ca are closer to the experimental values (red triangles). It is interesting to notice that, in the case of $^{40}$Ca, where the tensor force does not play any role, the better agreeement with the experimental gap is entirely due to the reduced strength of the spin--orbit contribution.   

\begin{figure}[htb]
\begin{center}
\includegraphics[width=8cm]{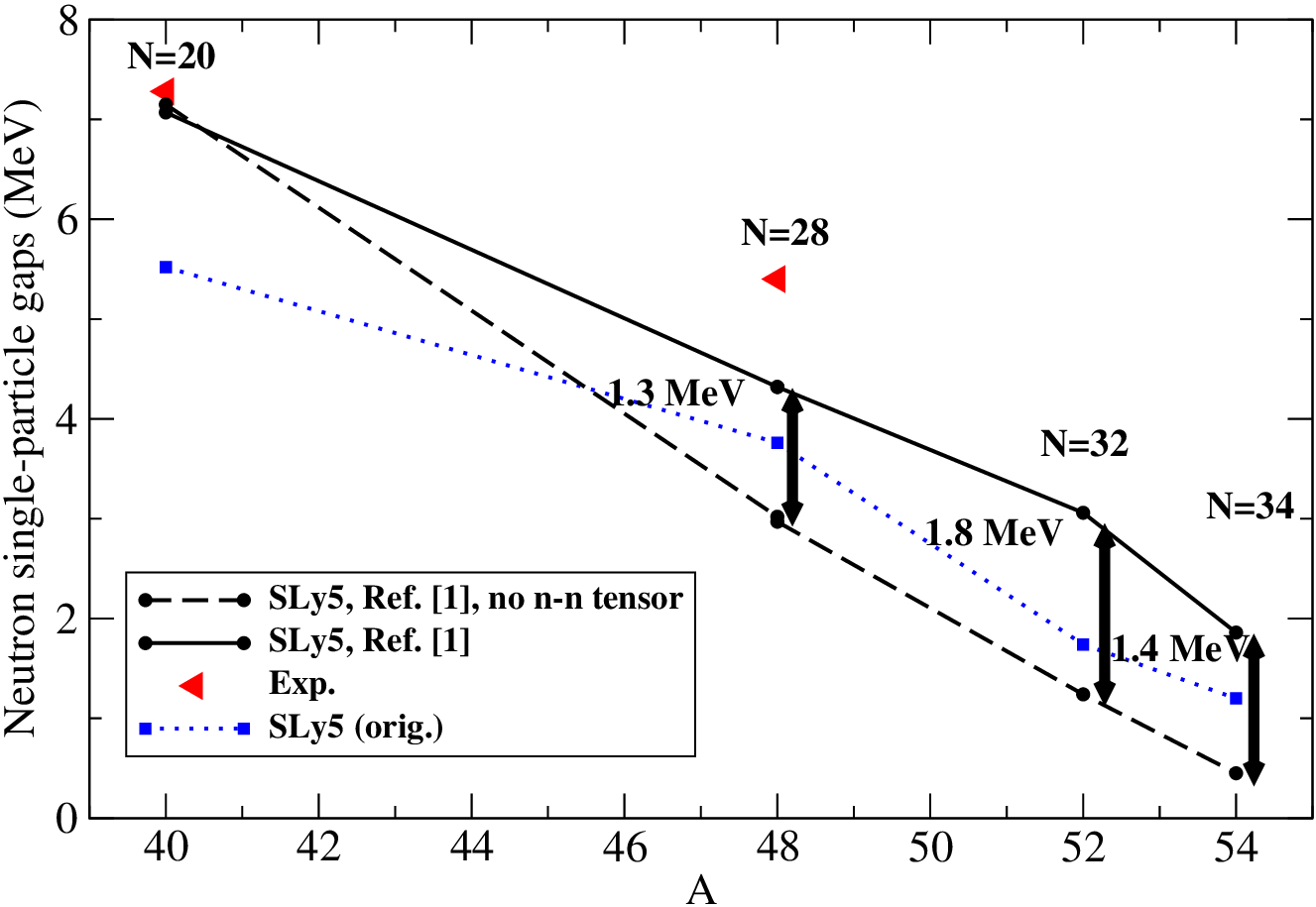}
\end{center}
\caption{(Color online) Single--particle neutron gaps for $^{40}$Ca, $^{48}$Ca, $^{52}$Ca, and $^{54}$Ca. The black solid line 
corresponds to the results obtained with the new Skyrme force; the black dashed line corresponds to the results obtained with the new Skyrme force by neglecting the n--n tensor contribution. 
The results obtained by using the original SLy5 force are also reported (blue dotted line). The experimental values for $^{40}$Ca and $^{48}$Ca are represented by red triangles.}
\end{figure}

Let us investigate in detail the tensor effect for each of the three isotopes where the tensor force  contributes. 
For $^{48}$Ca, the neutron $1f_{7/2}$ state is filled. Both high-$j$ (by using the 
terminology of Ref. \cite{otsuka}) single--particle states  $1f_{7/2}$ and $2p_{3/2}$ are pushed downwards, 
more strongly for the first state than for the second [Fig. 2(a)]. The net effect is that the gap between the two levels is increased. 

\begin{figure}[htb]
\begin{center}
\includegraphics[width=8cm]{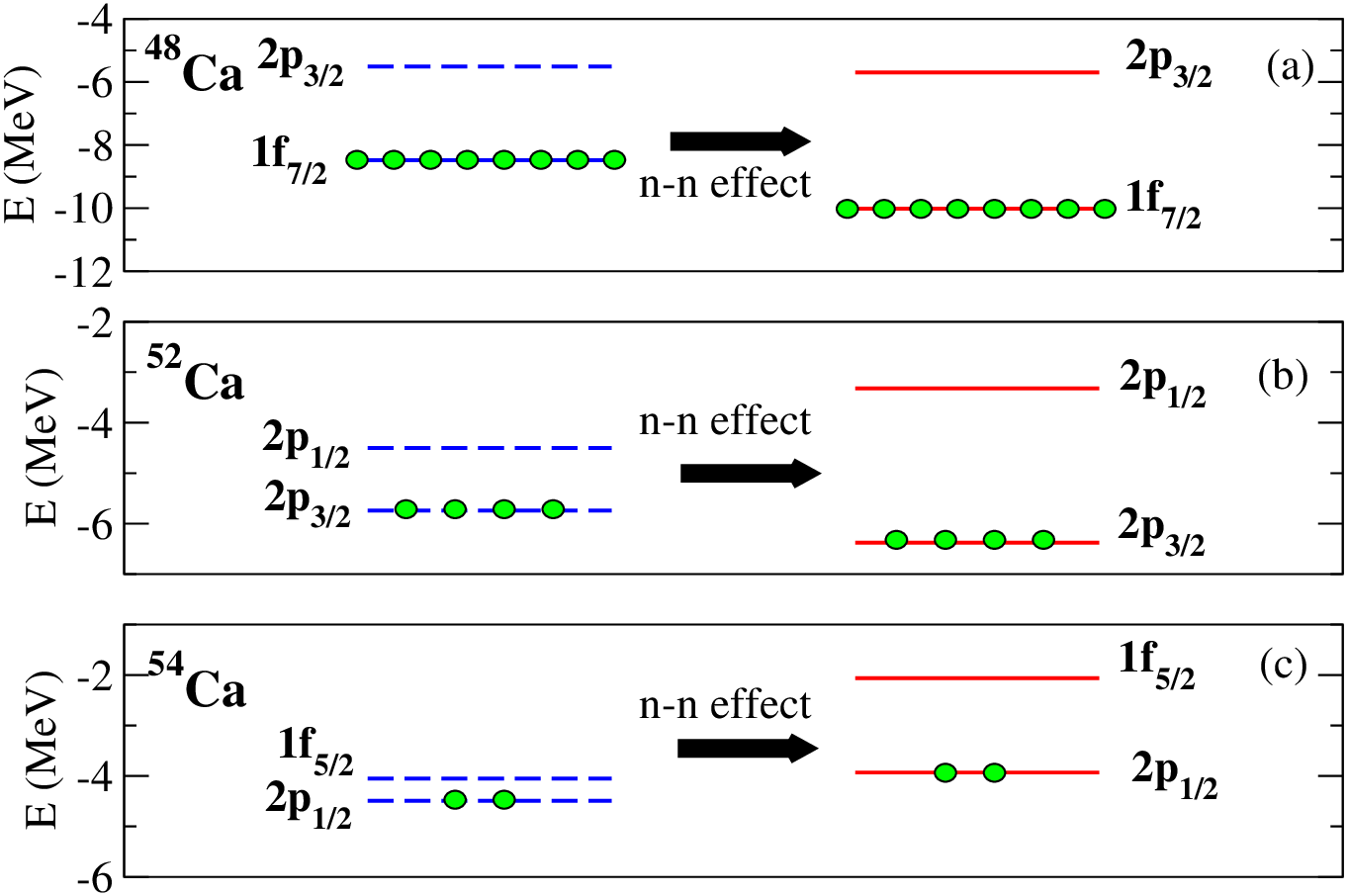}
\end{center}
\caption{(Color online) Neutron single--particle states for $^{48}$Ca (a), $^{52}$Ca (b), and $^{54}$Ca (c) 
with and without the like--particle tensor contribution.}
\end{figure}

For $^{52}$Ca, the neutron $2p_{3/2}$ state is filled. The high-$j$ $2p_{3/2}$ state is pushed downwards whereas the 
low-$j$ $2p_{1/2}$ state is pushed upwards [Fig. 2(b)]. The single--particle gap is increased. 

For $^{54}$Ca, the 
neutron $2p_{1/2}$ state is filled. Both low-$j$ single--particle states $1f_{5/2}$ and $2p_{1/2}$ are 
pushed upwards, more strongly for the first state than for the second [Fig. 2(c)]. The gap increases also this time. 

In particular, the stronger enhancement of the single--particle gap obtained in $^{52}$Ca is due to the 
fact that each single--particle state is pushed in the opposite direction by the tensor contribution.  
 Figs. 1 and 2 clearly show that 
the single--particle gaps are increased in $^{52}$Ca and $^{54}$Ca due to the like--particle tensor 
contribution. This provides an enhancement of magicity for the two isotopes, with respect to what found by neglecting the tensor term.  In particular, the gap is increased for $^{52}$Ca from 1.24 to 3.06 MeV and for $^{54}$Ca from 
0.45 to 1.86 MeV, providing in this way nuclei that have a stronger  
closed--shell nature. With the opposite sign of the tensor strength, the effect would be the opposite, that is, the gap 
would be decreased by including the tensor contribution and would be in particular more strongly shrinked for the nucleus 
$^{52}$Ca, where the $2p_{3/2}$ state would be shifted upwards and the $2p_{1/2}$ state would be pushed 
downwards. The presently used signs of the tensor parameters are thus the correct ones that allow us to enhance the magic character of the nuclei under study.

\section{Magicity of the isotopes $^{52}Ca$ and $^{54}Ca$. Neutron--proton tensor effects}
\label{np}

To analyze the effects of the n--p tensor contribution, isotonic chains containing $^{52}$Ca and $^{54}$Ca have to be analyzed. 

Let us start with the $N=32$ isotones $^{52}$Ca, $^{54}$Ti, $^{56}$Cr, and $^{58}$Fe. In these nuclei, the last occupied neutron state is the state $2p_{3/2}$, the upper level is the state $2p_{1/2}$, and the neutron gap $N=32$ thus coincides with the spin--orbit splitting of the $2p$ neutron states. Going from  $^{58}$Fe to $^{52}$Ca, the $Z$ number is reduced from 26 to 20. The occupation of the proton state $1f_{7/2}$ is thus decreased from 6 to 0. The n--p tensor effect is expected to provide an attractive interaction between the proton state $1f_{7/2}$ and the neutron state $2p_{1/2}$ and a repulsive interaction between the proton state $1f_{7/2}$ and the neutron state $2p_{3/2}$ in a given nucleus. 
Such tensor contribution is thus expected to induce a reduction of the neutron $p$ spin--orbit splitting, that is, a reduction of the $N=32$ gap. This effect is however expected to be weakened  along the isotonic chain, as far as the occupation of the proton $1f_{7/2}$ state is reduced, that is, going from $^{58}$Fe to $^{52}$Ca. 
With the reduction of the 
n--p effect, the gap is expected to increase going from $^{58}$Fe to $^{52}$Ca. This is shown in Fig. 3, where the gap is plotted (black solid line) and in Fig. 4, where the involved proton and neutron single--particle energies are displayed. 
In Fig. 4 one can see that, moving from the left to the right (that is, towards more neutron--rich nuclei along the isotonic chain), the energy of the proton state moves towards lower values and the energies of the neutron states are pushed upwards, as expected in mean--field calculations. On top of the global mean--field evolution, we should isolate the tensor contribution. 
From Fig. 3, one observes that  
the expected effect of enhancement of the gap going from $^{58}$Fe to $^{52}$Ca looks very weak. To better understand how the tensor force acts in this particular case, we compare the found results with those obtained by switching off the n--p tensor strength. The corresponding values are shown in Fig. 3 (red dashed line). One can observe that, without the n--p tensor contribution, the values are similar but the trend is the opposite, that is, the gap is reduced when passing from  $^{58}$Fe to $^{52}$Ca. The tensor force changes this trend. 
To isolate all the tensor contributions, we have repeated the same calculations by quenching also the parameter responsible for the like--particle tensor effect. The corresponding results are shown in the figure by a green dot-dashed line. It is clear that the like--particle tensor contribution provides an enhancement of the gap but the change of trend is due only to the n--p contribution. 
It is also interesting to compare the present results with the values obtained with the original SLy5 force (blue dotted line in Fig. 3). We can see that the trend is very similar to that obtained in our case by switching off the n--p tensor strength or both tensor strengths. There is however a shift of the values: the results obtained with the original parametrization are located between the green dot--dashed and the red dashed curves. The effect of the reduction of the spin--orbit induces the shift from the dotted to the dot--dashed curve and the inclusion of the like--particle tensor strength strongly pushes the values upwards. Finally, the inclusion of the n--p tensor contribution determines the change of slope and provides increasing (instead of decreasing) gaps going from $^{58}$Fe to $^{52}$Ca. This is what expected following the arguments based on shell--model calculations and presented in Ref. \cite{riken}, where this analysis is done for $N=34$ isotones. 

\begin{figure}[htb]
\begin{center}
\includegraphics[width=8cm]{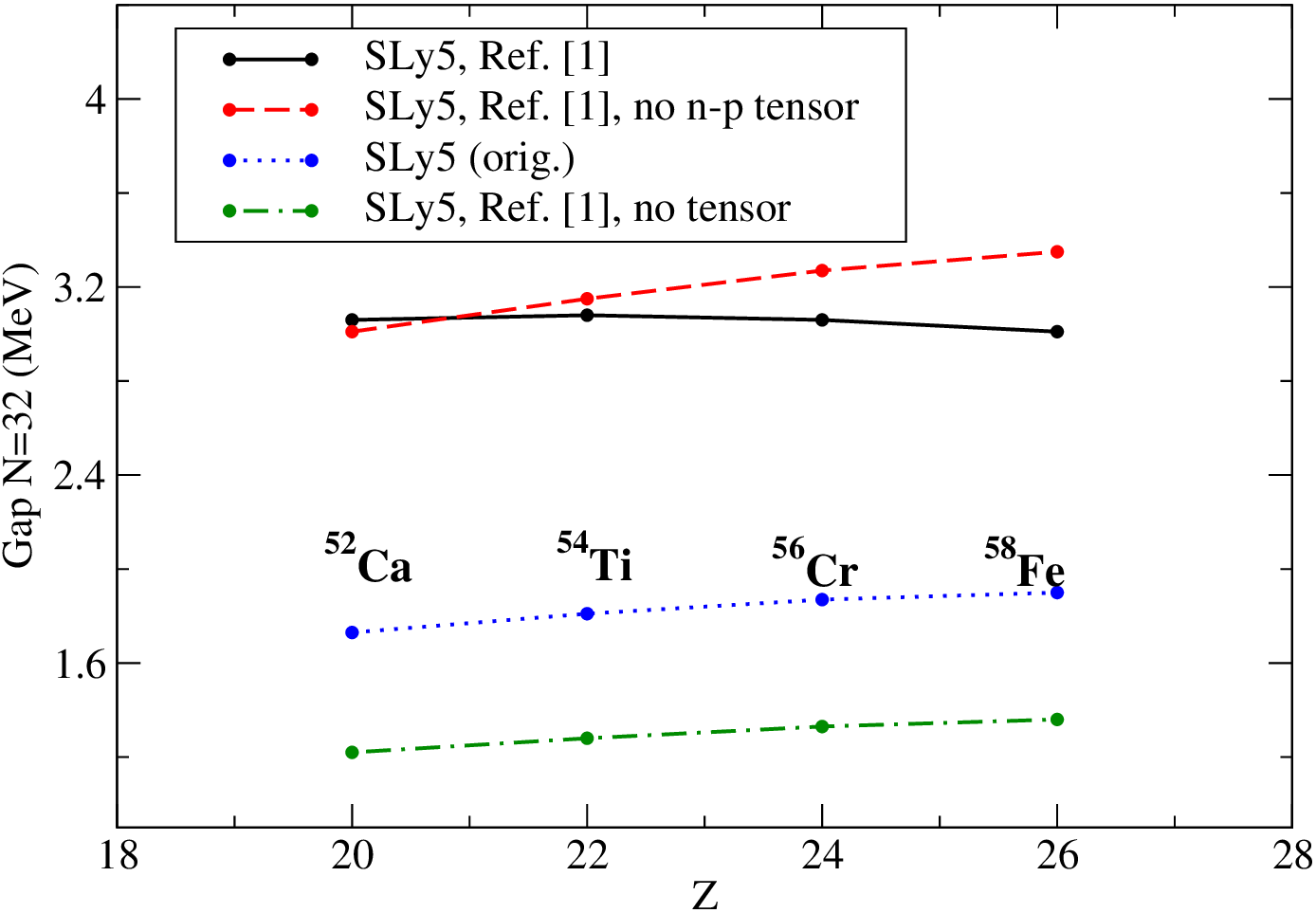}
\end{center}
\caption{(Color online) Neutron gap $N=32$ calculated for the isotones $^{52}$Ca, $^{54}$Ti, $^{56}$Cr, and $^{58}$Fe.}
\end{figure}

\begin{figure}[htb]
\begin{center}
\includegraphics[width=8cm]{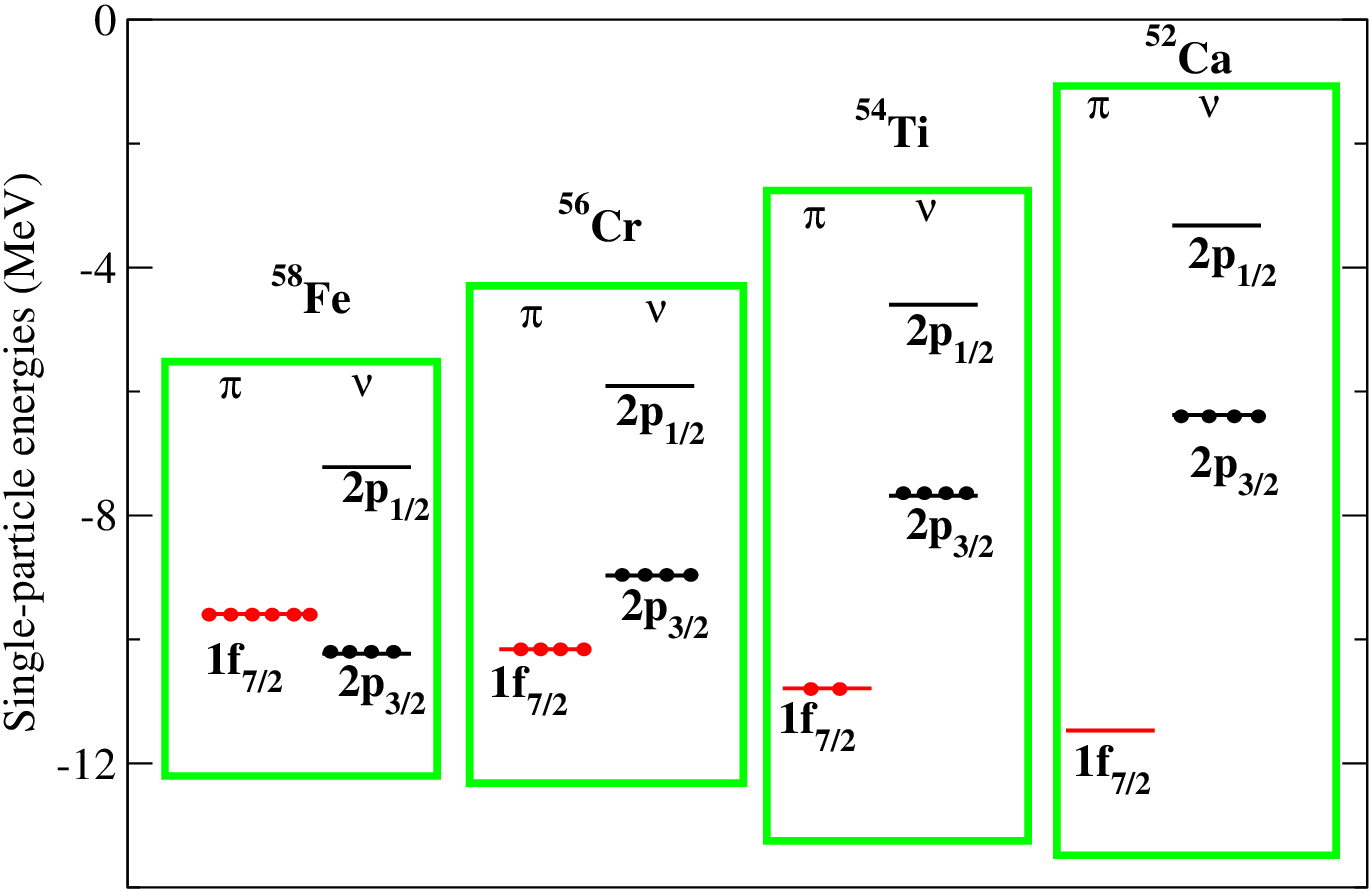}
\end{center}
\caption{(Color online) Single--particle energies of the proton state $1f_{7/2}$ and of the neutron states 
$2p_{3/2}$ and $2p_{1/2}$ for the isotones $^{52}$Ca, $^{54}$Ti, $^{56}$Cr, and $^{58}$Fe. }
\end{figure}

We repeat the same analysis for $N=34$ isotones for the nuclei 
$^{54}$Ca, $^{56}$Ti, $^{58}$Cr, and $^{60}$Fe.
The results are plotted in Figs. 5 and 6. The difference with respect to the previous case is that the last occupied neutron state is now the state $2p_{1/2}$ and the neutron gap $N=34$ is calculated between the states $1f_{5/2}$ and $2p_{1/2}$. Now, the n--p tensor effect associated to the proton state 
$1f_{7/2}$
induces for both neutron states an attractive interaction, stronger for the $f$ state than for the $p$ state. This means that the gap is expected also in this case to be reduced in a given nucleus owing to the n--p tensor contribution. Again, when this contribution is weakened going from $^{60}$Fe to $^{54}$Ca, the gap is eventually expected to increase. This effect may be seen in Fig. 5 (black solid line). 
In Fig. 6 the involved neutron and proton single--particle energies are represented. Again, the global mean--field evolution provides a reduction of the proton energy and  an enhancement of the neutron energies, when going towards more neutron--rich nuclei along the isotonic chain. On top of this general effect, the tensor effect has to be disentangled. This is done by plotting in Fig. 5 the results obtained by switching off the n--p tensor strength (red dashed line). The gap still increases, but much less strongly than in the full calculations. 
By switching off also the like--particle tensor strength, the obtained values are reported by a green dot-dashed line. The value for the nucleus $^{60}$Fe is not reported. The Hartree--Fock calculation does not converge in this case. This is probably due to the fact that the neutron states $2p_{1/2}$ and $1f_{5/2}$ become almost degenerate and cross each other at each iteration. 
The values obtained with the original SLy5 force are also presented (blue dotted line). Also this time one observes that the slope in the results obtained with the original SLy5 force  is very similar to that obtained by switching off, in our case, the n--p tensor strength or both tensor  strenghts. Also in this case there is a shift between the dotted curve and the dashed and dot--dashed curves. 
The spin--orbit reduction in the new parametrization determines the shift to lower values from the dotted to the dot--dashed curve. The inclusion of the like--particle tensor contribution induces a strong shift to higher energies (dashed curve). Finally, the inclusion of the n--p tensor contribution leads to a change of slope (solid curve). 
We notice that this time, even without the inclusion of the n--p tensor contribution, the mean--field calculations provide an enhancement of the gap when going from $^{60}$Fe to $^{54}$Ca. The tensor force leads however to a much stronger effect, affecting significantly the slope.  

As was mentioned in Sec. I, the calculations are done in spherical symmetry and the effect of possible deformations is thus neglected. The inclusion of possible small deformations for some $N=32$ and $N=34$ isotones would probably modify quantitatively our predictions but we do not expect that the qualitative trends and the general conclusions of this work would be changed: the tensor force would always shift in the same direction the single--particle energies; the qualitative evolution of the corresponding gaps would not be expected to be stronlgy modified.  

\begin{figure}[htb]
\begin{center}
\includegraphics[width=8cm]{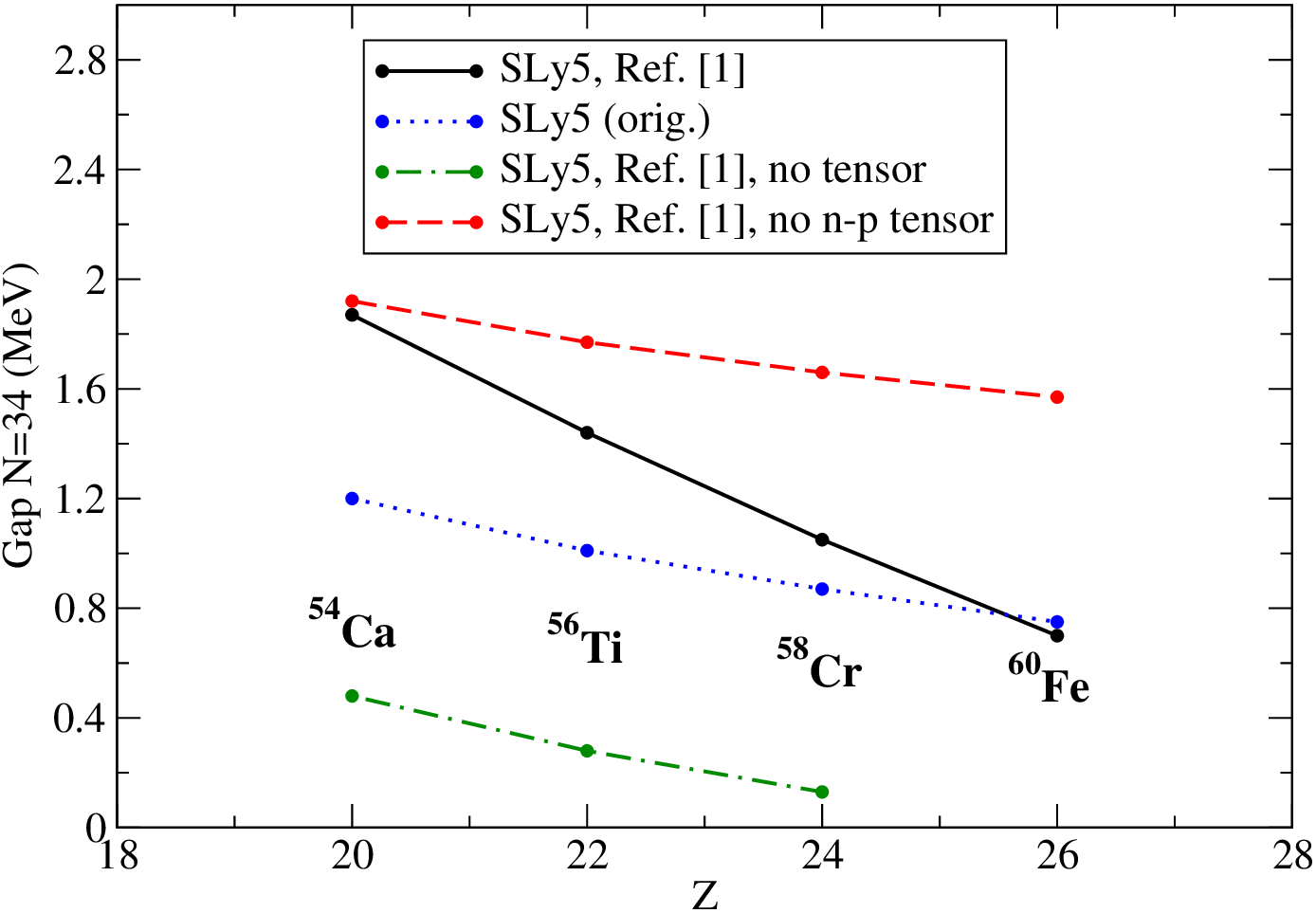}
\end{center}
\caption{(Color online) Neutron gap $N=34$ calculated for the isotones $^{54}$Ca, $^{56}$Ti, $^{58}$Cr, and $^{60}$Fe. }
\end{figure}

\begin{figure}[htb]
\begin{center}
\includegraphics[width=8cm]{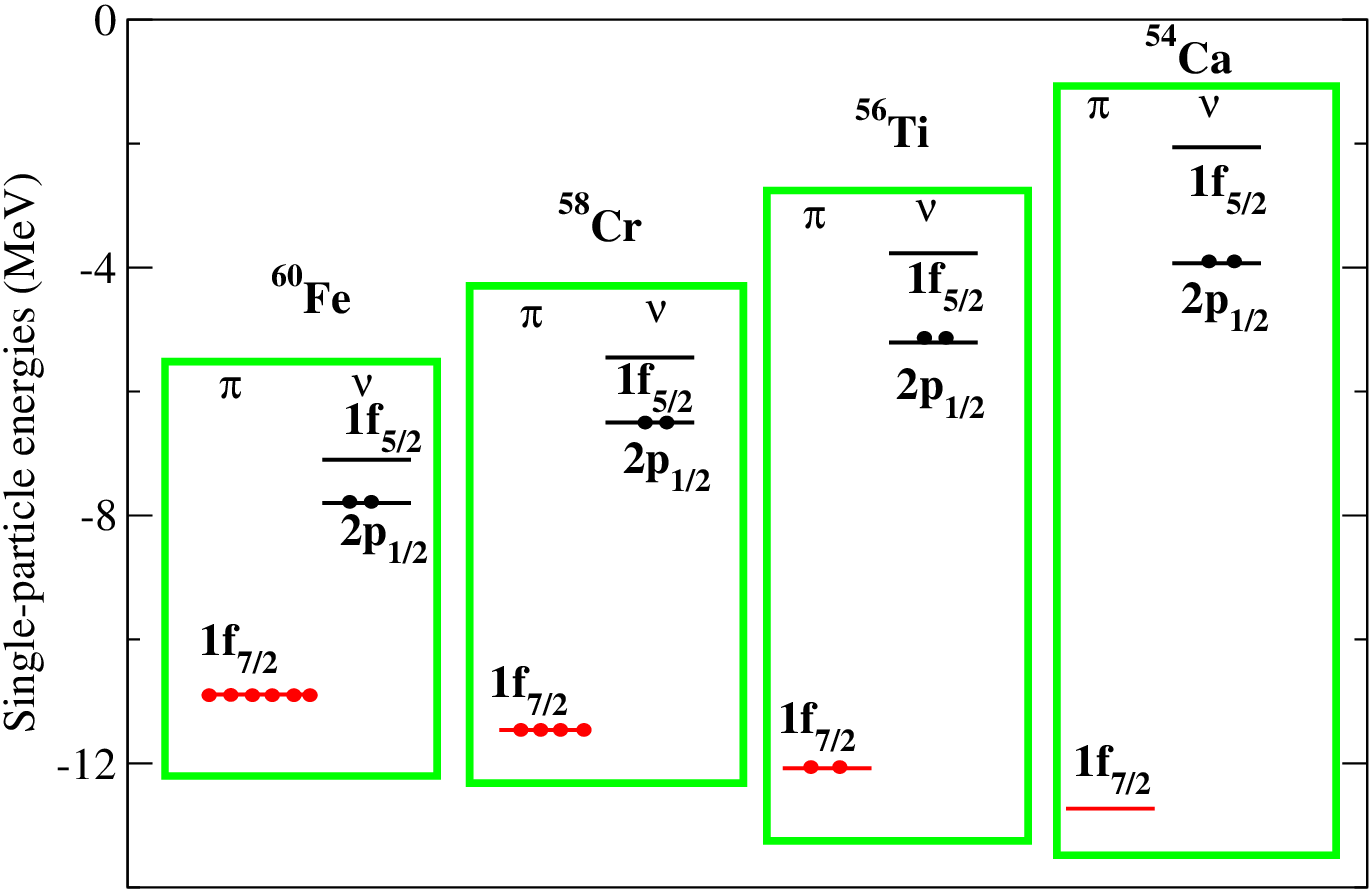}
\end{center}
\caption{(Color online) Single--particle energies of the proton state $1f_{7/2}$ and of the neutron states 
$2p_{1/2}$ and $1f_{5/2}$ for the istones $^{54}$Ca, $^{56}$Ti, $^{58}$Cr, and $^{60}$Fe.}
\end{figure}

\section{Masses and separation energies}
\label{masses}
It was stressed in Ref. \cite{grasso2013}  that the new sets may induce non negligeable
effects on the masses of some nuclei, because the spin--orbit strength was modified without refitting also the other parameters of the force. We have thus checked whether the 
values of the masses predicted for the Ca isotopes under study are reasonable with the new 
parametrization. We have found the following binding energies: 343.83, 415.91, 438.62, and 445.21
MeV for the nuclei $^{40}$Ca, $^{48}$Ca, $^{52}$Ca, and 
$^{54}$C, respectively. These values have to be compared with the binding energies obtained with the original SLy5 
force, that are 344.07,  415.92, 437.25, 
and 444.94 MeV for $^{40}$Ca, $^{48}$Ca, $^{52}$Ca, 
and $^{54}$C, respectively. 
We observe that the deviations are not very important. The largest deviation between the new Skyrme  values and those 
obtained with SLy5 is 0.3\%. 

In Ref. \cite{wienholtz}, the two--neutron separation energies are compared with several theoretical models. 
Experimentally, a clear signature of shell closure for the nucleus $^{52}$Ca is the important 
reduction of the $S_{2n}$ value going from $^{52}$Ca to $^{54}$Ca. This new experimental value is now available owing  to the high--precision mass measurement of 
$^{54}$Ca reported in Ref. \cite{wienholtz}: 
going from $^{50}$Ca to $^{52}$Ca, the value of $S_{2n}$ remains almost the same, and it has a sudden decrease from $A=52$ to $A=54$.
Several EDF results are reported in Fig. 3(b) of 
Ref. \cite{wienholtz}. It is commented there that, in general, these models do not provide the correct trend for the 
$S_{2n}$ energies going from $A=50$ to $A=54$: they predict a very smooth (almost linear) change in the $S_{2n}$ values.  

We have thus computed the two--neutron separation energies 
$S_{2n}= E(N,Z)-E(N-2,Z)$, where $E(N,Z)$ is the binding energy of the nucleus 
$(N,Z)$,
and compared them with the experimental values for the isotopes  
$^{50}$Ca, $^{52}$Ca, and 
$^{54}$Ca. This is shown in Fig. 7 where also the experimental values (including those of Ref. \cite{wang} and the new data of Ref. \cite{wienholtz}) are presented. 

\begin{figure}[htb]
\begin{center}
\includegraphics[width=8cm]{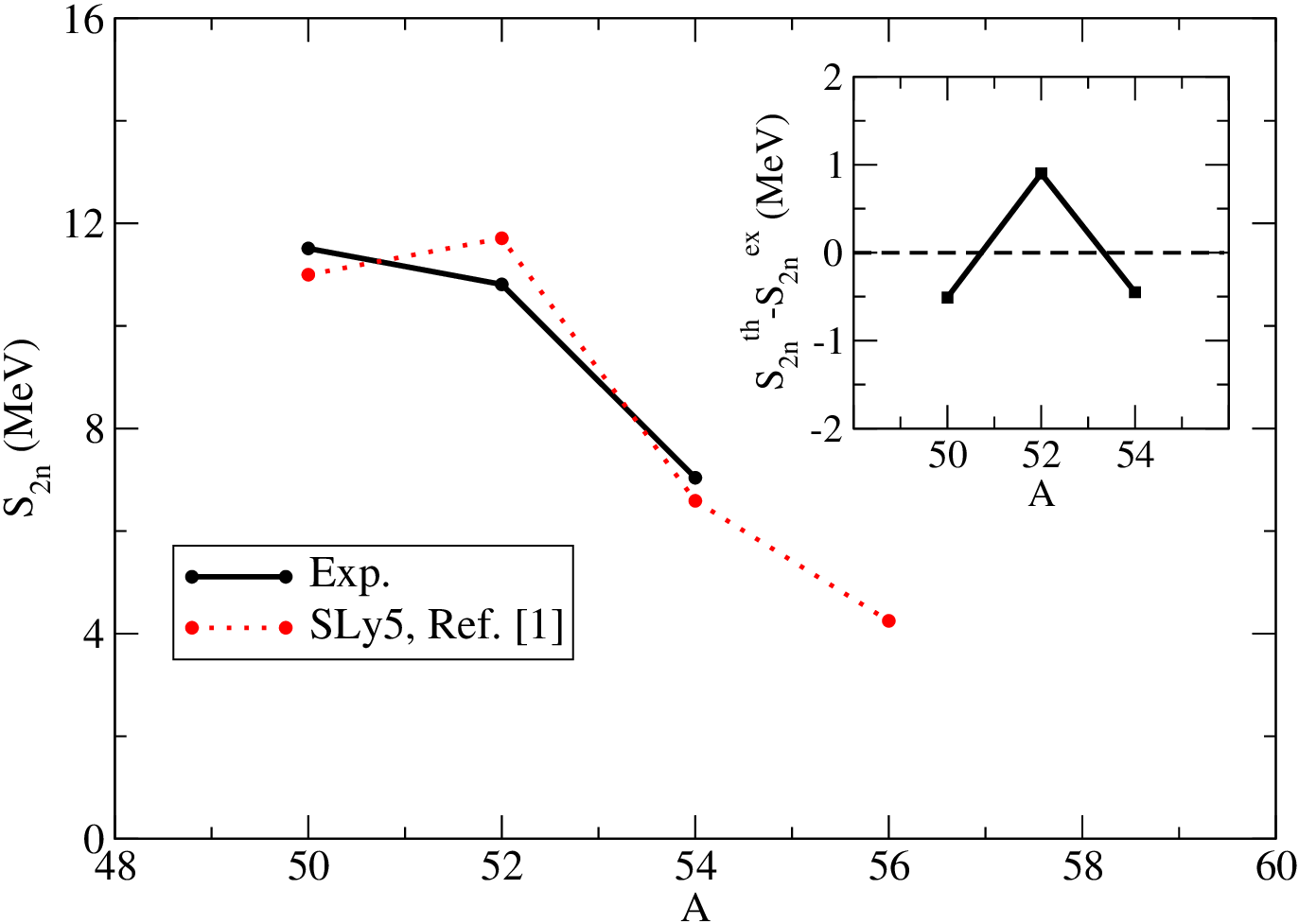}
\end{center}
\caption{(Color online) Two--neutron separation energies for the isotopes $^{50}$Ca, $^{52}$Ca, and 
$^{54}$Ca calculated with the new Skyrme set. The experimental values are reported \cite{wienholtz,wang}. The theoretical value for the nucleus $^{56}$Ca is also shown. In the inset the difference between the theoretical and the experimental values is displayed.}
\end{figure}

We
observe that the agreement with the experimental values is quite good and that the predicted change with respect to $A$ is not linear. The good agreement can also be seen in the inset of the figure, where the 
differences between the theoretical and the experimental values of $S_{2n}$ are reported. We have to mention  that the correct trend for the $S_{2n}$ values may be found also by using the original SLy5 force, without including the tensor term. However, the signatures of magicity are slightly  stronger with the present set of parameters. With the present modified SLy5, the so--called two--neutron shell gap, calculated as $S_{2n}(N,Z)-S_{2n}(N+2,Z)$, is equal to 5.1 MeV for $^{52}$Ca, to be compared with an experimental value of almost 4 MeV \cite{wienholtz}. 

The reduction of the $S_{2n}$ value, found going from $A=54$ to $A=56$,  cannot be compared with any experimental result because the mass of the nucleus 
$^{56}$Ca was not yet measured experimentally. The experimental $S_{2n}$ value is thus still unknown at $A=56$.  
However, our theoretical prediction, 
that leads to a quite significant reduction from $A=54$ to $A=56$, 
 is coherent with the new spectroscopic data of Ref. \cite{riken}, that indicate a magic nature for the nucleus $^{54}$Ca. 
One should mention however that the inclusion of possible pairing correlations in the nucleus $^{56}$Ca would render this system more bound and would thus provide a higher value of $S_{2n}$ at $A=56$.

We can conclude that the present EDF results display the correct trend for the values of $S_{2n}$ in the region from 
$^{50}$Ca to $^{54}$Ca (weak change from $A=50$ to $A=52$ and significant change from $A=52$ to $A=54$).

\section{Conclusions}
\label{concl}

In this work, we have examined some shell effects that may be clearly related to the contributions induced by a 
tensor force within the mean--field framework with the Skyrme interaction. The employed 
tensor force is of zero range. We use one parametrization that was 
introduced in a previous work \cite{grasso2013} on top of the Skyrme 
SLy5 force \cite{sly5}. 
In this parameter set, the spin--orbit strength was adjusted together with the tensor parameters. With this parameter set, we have performed Hartree--Fock calculations in spherical symmetry. We are aware that many effects are disregarded by using this simple theoretical model, but our objective is to isolate the genuine effects coming from the inclusion of the tensor force in a mean--field model.  

The magicity of the two Ca isotopes $^{52}$Ca and $^{54}$Ca has been confirmed by two recent experimental 
measurements \cite{wienholtz,riken}. We have shown here that the introduction of the tensor force leads to 
an enhancement of magicity in the two isotopes.
By analyzing Ca isotopes, it is shown that the like--particle tensor contribution 
  renders these nuclei more magic than they would be predicted by neglecting it. 
By studying $N=32$ and $N=34$ isotones, the neutron--proton tensor effects are identified and isolated with respect to the effects coming from the like--particle tensor contribution. It is shown that the $N=32$ and $N=34$ neutron gaps are predicted to increase along isotonic chains, going from $^{58}$Fe to $^{52}$Ca, and from $^{60}$Fe to $^{54}$Ca, respectively.  
In the case $N=32$, the n--p tensor contribution is responsible for the increasing trend (without this contribution, the gap decreases). In the case $N=34$, the gap increases even without the n--p tensor contribution, but the n--p tensor contribution  determines an important change of the slope. 

A check on the masses of the Ca isotopes under study is performed and the two--neutron separation energies are compared with the experimental data for $^{50}$Ca, $^{52}$Ca, and $^{54}$Ca. A good agreement is found: from $A=50$ to $A=52$, the $S_{2n}$ value does not change strongly, whereas it is significantly reduced from $A=52$ to $A=54$. This provides an evidence for the shell closure $N=32$  in Ca isotopes.     

The application described in this work confirms the robustness of the findings of Ref. \cite{grasso2013} about the signs 
and the  values for the tensor parameters. The employed parameter set, even if introduced with a partial adjustment of the parameters (the other parameters of the forces 
are not modified, with the exception of the spin--orbit strength), provides the correct expected shell 
effects. 
Different signs of the tensor parameters would generate un uncorrect behavior if compared with the experimental results. 
The present tensor parametrization (and its induced like--particle and neutron--proton effects) allows us in particular to describe, within a simple mean--field 
picture, the magicity of the Ca isotopes $^{52}$Ca and $^{54}$Ca.

%
%
%
%-----------------------------------

\end{document}